\documentclass{article}
\usepackage{graphicx}
\usepackage{authblk}
\usepackage{arxiv}
\usepackage{amsmath}
\usepackage[utf8]{inputenc} 
\usepackage{amsfonts}       
\usepackage{caption}

\begin{document}
\title{Flip and Neimark-Sacker Bifurcations in a Coupled Logistic Map System}

\author{A. Mareno$^1$ and L.Q. English$^2$}
\affil{$^1$Department of Mathematics and Computer Science, Pennsylvania State University, Capital College, Middletown PA, 17057 \\ $^1$email: aum24@psu.edu\\ $^2$Department of Physics, Dickinson College, Carlisle, PA 17013}

\maketitle
\begin{abstract}
In this paper we consider a system of strongly coupled logistic maps involving two parameters. We classify and investigate the stability of its fixed points.  A local bifurcation analysis of the system  using Center Manifold  is undertaken and then supported by numerical computations.This reveals the existence of reverse flip and Neimark-Sacker bifurcations.
\end{abstract}


\section{Introduction}
Coupled logistic maps originally gained attention in the mathematical biology literature via their utility in models of, for instance, populations of migrating species and environmental heterogeneity \cite{gyllenberg,kendall}. Recent years, however, have seen a renewed interest in the dynamics of coupled logistic maps. At least two developments have spurred this re-examination: (a) the realization that discrete coupled maps could be usefully exploited in digital encryption schemes \cite{solis,askar,elsadany}, and (b) success with their experimental implementation using electronic circuits \cite{lher,mhiri,english}. Both of these recent threads have revealed intricate and non-intuitive behavior of these coupled maps. One such behavior – spontaneous symmetry breaking - was recently highlighted and explored \cite{english}. That reference, however, did not attempt to analyze the chaotic regime in this coupled system (i.e., for large values of $r$), focusing primarily on symmetry breaking and its basin of attraction pertaining to n-cycles. Here we revisit the problem in a mathematically rigorous way and thus shed light on the origins of some of the unusual bifurcations seen experimentally \cite{mhiri} in this system.
 
In particular, we start by systematically classifying all the fixed points and their bifurcation properties that manifest in this coupled system, taking the coupling strength, $\epsilon$, to be our bifurcation parameter and not  the growth rate, $r$, which is typically chosen. We then focus on the symmetry-broken 1-cycle – a fixed point unique to the coupled system -  and proceed to apply the center-manifold-theoretic framework to prove that it becomes stable via a flip bifurcation as the coupling strength parameter is increased. This transition is an interesting phenomenon also seen previously in experiments \cite{mhiri}. In this paper, we explore this flip bifurcation also numerically and see excellent agreement with the predictions derived from theorem established herein.

For even higher values of the coupling strength, the symmetry-broken 1-cycle loses stability again (something also seen experimentally). In this context, we prove that the origin of this instability is a Neimark-Sacker bifurcation. We then explore this bifurcation numerically, and again demonstrate excellent agreement with our theoretical results.

Throughout this work we consider the following discrete system
\begin{eqnarray}
x_{n+1}=(1-\epsilon) f(x_n) + \epsilon f(y_n) \nonumber \\
y_{n+1}=\epsilon f(x_n) + (1-\epsilon) f(y_n),
\label{full}
\end{eqnarray}
where 
\begin{equation}
f(z)=rz(1-z).
\end{equation}
For convenience the system can be rewritten in the form:
\begin{equation}
    F(x,y)=((1-\epsilon)f(x)+\epsilon f(y), \epsilon f(x)+(1-\epsilon)f(y)) \label{main}
\end{equation}
where the parameter $\epsilon \in [0,1]$ and $r\in(0,4).$

The organizational structure of this paper is as follows:  In section 2 we discuss the basic framework and relevant terminology for this work. In section 3, we classify and determine the stability of the fixed points of the system the $r\epsilon$ plane. Section 4 is devoted to a rigorous mathematical treatment of the flip and Neimark-Sacker bifurcations manifested by this system. Finally, in section 5 we give numerical evidence to support our theoretical results from section 4.


\section{Invariant Manifolds and Center Manifold Theory}
We begin by stating important terminology and concepts relevant to this work (see for instance \cite{CA, WG}). Generally we can say that a set $S$ is an invariant set if iterates of the map for any element of S stay in S for all integers. We will loosely think of an invariant manifold as a set which locally has the structure of Euclidean space, typically as surfaces imbedded in $\mathbb{R}^n$, for which the function representing the surface has maximal rank and can therefore, be locally represented as a graph, by way of applying the Implicit Function Theorem. \\[0.1in]
We now define three important linear subspaces, relevant to the study of dynamical systems, spanned by the (generalized)
eigenvectors of the Jacobian matrix $DF(x,y)$ at a fixed point $(x,y)$:  $E^{s}$(the stable subspace), $E^{u}$(the unstable subspace) and $E^{c}$(the center subspace). The associated eigenvalues of each subspace have modulus less than one, greater than one or equal to one respectively. 
When $DF(x,y)$ has no eigenvalues of unit modulus $(x,y)$ is called a {\it hyperbolic point} and so its stability is determined entirely by the eigenvalues themselves.  Furthermore, for hyperbolic points $E^{c}$ does not exist. \\[0.1in]
A hyperbolic fixed point is called a {\it sink} if the eigenvalues of the Jacobian matrix evaluated at the fixed point have magnitude less than one.  Such a fixed point is locally asymptotically stable. If the magnitudes of both eigenvalues are greater than one, the hyperbolic fixed point is called a {\it source} and is locally asymptotically unstable.  Moreover, a hyperbolic fixed point is called a  {\it saddle point} if only one of the eigenvalues has magnitude greater than one. \\[0.1in]
The Stable Manifold Theorem \cite{GH} guarantees the existence of local stable and unstable invariant manifolds $W^{s}_{loc}$ and $W^{u}_{loc}$ which can be viewed as nonlinear analogues of the linear subspaces $E^{s}$ and $E^{u}$ respectively.  These invariant manifolds are tangent to these the two linear subspaces, have the same dimensions as these subspaces and are as smooth as the underlying map.  \\[0.1in]
The Center Manifold Theorem ( see chapter 1 in \cite{GH} or \cite{CA} )asserts the existence of an invariant manifold tangent to the center eigenspace $E^{c}$ which can be non-unique and `non-smooth' (in a certain sense) (see-chapter 3 in \cite{GH} or\cite{CA}) where the dynamics of the nonlinear system (at say the trivial fixed point) restricted to the center manifold is determined by a c-dimensional map, a map whose dimension is the same as that of the center subspace $E^{c}$, where say $(x,y)\in{\mathbb{R}^c \times \mathbb{R}^s}$ and both $\mathbb{R}^s,\mathbb{R}^c$ are subsets of $\mathbb{R}^n$. So for a two-dimensional system such as the system studied in this paper the dynamics of our nonlinear map are determined by a one-dimensional map. \\[0.1in]
Herein lies the significance of Center Manifold Theorem - rather than studying the map on the entire domain of the map to determine its dynamics we can restrict this analysis to the center manifold, an invariant manifold with dimension equal to the dimension of the center subspace, which is less than the dimension of the maps' domain. In addition, using the invariance of the center manifold one can derive a quasi-linear partial differential equation that the c-dimensional map characterizing the center manifold must satisfy in order for its graph to be an invariant center manifold. To find this map, one must solve this partial differential equation. Thus this theorem can be viewed as type of reduction principle that one can apply to ascertain the stability of non-hyperbolic fixed points, when say $E^{u}$ is trivial.\\[0.1in]
Therefore, in this paper we restrict our use of Center Manifold Theory to the case where the Jacobian matrix has its spectrum inside the unit circle apart from one or two eigenvalues. For an additional reference on Center Manifold Theory see \cite{ELY}.  \\[0.1in] 
\section{Classification of the fixed points of the nonlinear system }
We begin our analysis of the system ~(\ref{main}) by solving the equations
\begin{align}
    (1-\epsilon)rx(1-x)+\epsilon ry(1-y)=x\\
    \epsilon rx(1-x)+(1-\epsilon)ry(1-y)=y.
\end{align}
and obtaining the fixed points of our system: 
\begin{equation}
    (0,0), (\frac{r}{r-1},\frac{r}{r-1}), (x*,y*), (y*,x*)
    \end{equation}
    where 
\begin{equation}
x^*=\frac{r(2\epsilon-1)+1-\sqrt{(r(1-2\epsilon)-1)(r(1-2\epsilon)+4\epsilon-1)}}{2r(2\epsilon-1)}
\label{line1}
\end{equation}
\begin{equation}
y^*=\frac{r(2\epsilon-1)+1+\sqrt{(r(1-2\epsilon)-1)(r(1-2\epsilon)+4\epsilon-1)}}{2r(2\epsilon-1)}.
\label{line2}
\end{equation}
We note that $x^{*},y^{*}$ are real valued if and only if $\Delta=(1-4\epsilon)(r-1)^2+4\epsilon^2r(r-2)\geq 0.\mbox{ This occurs when }$\\
\begin{equation}
    r\in(3,4) \mbox{ and } \epsilon \in\left[0,\frac{r-1}{2r}\right] \mbox{ or } \epsilon\in\left[\frac{r-1}{2(r-2)},1\right]
\end{equation}
In addition $x{*}={y*}$ if and only if $\Delta=0$ which occurs when $\epsilon=\frac{r-1}{2r}$ or $\epsilon=\frac{r-1}{2(r-2)}$, and so for these values of $\epsilon$ the fixed point $(x*,y*)$ coincides with one of the two symmetric fixed points: $(0,0)$ or $(\frac{r-1}{r},\frac{r-1}{r})$, respectively. Throughout this work we consider only $(x*,y*)$ and not $(y*,x*)$-its flipped counterpart. 
To determine conditions for a fixed point to be classified as a hyperbolic/non-hyperbolic fixed point and to determine the stability type of hyperbolic fixed points we compute the Jacobian of our map $F$ : \\

\begin{equation}
DF(x,y)=
\left(\begin{array}{cc} 
(1-\epsilon)r(1-2x) & \epsilon r(1-2y) \\
\epsilon r(1-2x) &  (1-\epsilon)(1-2y)
\end{array} \right)
\end{equation}
By solving the characteristic equation 
\begin{equation}
det(DF(x,y)-\lambda I)=0,
\end{equation}
the eigenvalues of the Jacobian evaluated at a fixed point $(x,y)$ are computed as follows:
\begin{equation}
\lambda_{1,2}=r(1-\epsilon)+r(\epsilon-1)(x+y)\pm\sqrt{r^2(1-2\epsilon)(x-y)^2+ \epsilon^2(x+y-1)^2} \label{eigen}
\end{equation}
Although the characteristic equation is characterized by the three principle invariants, where each is in turn is a function of the eigenvalues of the Jacobian and therefore one can use say, the Jury conditions to determine the stability of the fixed points, we take a more straight forward approach and analyze the eigenvalues and their magnitudes directly; this direct approach yields more `directional' information about the magnitudes of both eigenvalues.


Using these definitions and the eigenvalues associated with each fixed point we  determine the parameter dependent regions where each of the fixed points is asymptotically stable, unstable, a saddle point and a non-hyperbolic point, which are stated in the following theorem: \\[0.1in]
{\bf Theorem 1-Fixed Point Classification and Stability }\\[0.1in]
 A.  (i)The fixed point $(0,0)$ is sink if $r\in(0,1)$ and $\epsilon\in[0,1]$.\\[0.1in]
     (ii) (0,0) is a source if $r\in(0,1)$ and $\epsilon\in[0,\frac{r-1}{2r})$  or $\epsilon\in(\frac{r+1}{2r},1]$. \\[0.1in]
     (iii) (0,0) is a saddle point $r\in(1,4)$ and $\epsilon\in\left(\frac{r-1}{2r},\frac{r+1}{2r}\right)$. (Here, $|\lambda_1|>1 \mbox{ and } |\lambda_2|<1$).\\[0.1in]
     (iv) (0,0) is a non-hyperbolic point (specifically here, $\lambda_2=-1 \mbox{ and } |\lambda_1|>1)$ if $$ r\in(1,4) \mbox{ and } \epsilon=\frac{r+1}{2r}$$
     $$\lambda_2=1, |\lambda_1|>1 \mbox{ for }r\in(1,4), \epsilon=\frac{r-1}{2r} $$
     $$\lambda_1=1, \lambda_2=-1 \mbox{ for } \epsilon=1, r=1 $$
     $$\lambda_1=1, |\lambda_2|<1 \mbox{ for } \epsilon\in(0,1),r=1 $$
     $$\lambda_1=\lambda_2=1 \mbox{ for } \epsilon=0,r=1\mbox{ } \mbox{(1:1 resonance)}.$$
 B.  (i) The symmetric fixed point $(\frac{r-1}{r}, \frac{r-1}{r})$ is a sink if $$r\in(1,3)\mbox{ for all } \epsilon\mbox{ in } [0,1].$$\\
     (ii) $(\frac{r-1}{r}, \frac{r-1}{r})$ is a source if $$r\in(0,1)\mbox{ and } \epsilon\in\left[0, \frac{r-1}{2(r-2)}\right)\mbox{, or } r\in (0,1) \mbox{ and }\epsilon\in\left( \frac{r-3}{2(r-2)},1\right]$$\\
     or \\
     $$r\in(3,4)\mbox{ and } \epsilon\in\left[0, \frac{r-3}{2(r-2)}\right) \mbox{, or } r\in(3,4) \mbox{ and } \epsilon\in\left( \frac{r-1}{2(r-2)},1\right].$$\\
    
(iii) $(\frac{r-1}{r}, \frac{r-1}{r})$ is a saddle point (in this case it means $|\lambda_1|<1$ \mbox{and } $|\lambda_2|>1$) if $$r\in(0,1) \mbox{ and } \epsilon\in\left(\frac{r-1}{2(r-2)},\frac{r-3}{2(r-2)}\right)$$ \mbox{or}\\
$$r\in(3,4)\mbox{ and } \epsilon\in\left(\frac{r-3}{2(r-2)},\frac{r-1}{2(r-2)}\right).$$\\
(iv)$(\frac{r-1}{r},\frac{r-1}{r})$ is a non-hyperbolic point  \mbox{ if }  $$r\in(0,1) \mbox{ or } r\in(3,4) \mbox{ and } \epsilon=\frac{r-3}{2(r-2)} \mbox{ (here }\lambda_1=-1, |\lambda_2|> 1)$$  \mbox{ or }
 $$\epsilon\in(0,1) \mbox{ and } r=3 \mbox{ (here } \lambda_2=-1, |\lambda_1|< 1 )$$ \mbox{ or}
 $$\epsilon\in(0,1)\mbox{ and } r=1 \mbox{ (here } \lambda_2=1, |\lambda_1|< 1 )$$ \mbox{ or}
 $$\epsilon=\frac{r-1}{2(r-2)} \mbox{ and } r\in(0,1) \mbox{ or } r\in(3,4) \mbox{ and } \lambda_1=1, |\lambda_2|> 1 ).$$
 Furthermore, $$ \lambda_1=\lambda_2=-1, \epsilon=0,r=3 \mbox{ (1:2 resonance )}$$
 $$\lambda_1=1,\lambda_2=-1, \epsilon=1, r=3.$$
C.  (i) The non-symmetric fixed point $(x*,y*)$ is a sink if 
$$r\in(3,1+\sqrt{6}) \mbox{ and } \epsilon\in\left(\frac{1}{2}+\frac{\sqrt{3}}{2}\sqrt{\frac{1}{r(r-2)}},1\right] \mbox{ or } r\in(1+\sqrt{6},4) \mbox{ and } \epsilon\in\left(\frac{1}{2}+\frac{\sqrt{3}}{2}\sqrt{\frac{1}{r(r-2)}}, f_{2}(r)\right)$$ where $$f_{2}(r)= \frac{1}{4}\left[ \frac{3-4r+2r^2}{r(r-2)}+\frac{\sqrt{9-16r+8r^2}}{r^2(r-2)^2}\right]. $$\\
(ii)$(x*,y*)$ is a source if $$r\in(3,4)\mbox{ and } \epsilon\in\left[0,\frac{1}{2}-\frac{\sqrt{3}}{2}\sqrt{\frac{1}{r(r-2)}}\right)$$ \mbox{ or }\\
$$ r\in(1+\sqrt{6},4)\mbox{ and } \epsilon\in(f_2(r),1].$$\\
(iii) $(x*,y*)$ is a saddle point if $r\in(3,4)$ and  
$$ \epsilon\in\left(\frac{r-1}{2(r-2),}\frac{1}{2}+\frac{\sqrt{3}}{2}\sqrt{\frac{1}{r(r-2)}}\right], \mbox{ (specifically } |\lambda_2|<1, |\lambda_1|>1)$$\\
\mbox{ or }
$$ r\in[3,4) \mbox{ and } \epsilon\in\left(\frac{1}{2}-\frac{\sqrt{3}}{2}\sqrt{\frac{1}{r(r-2)}},\frac{r-1}{2r}\right),\mbox{ (specifically } |\lambda_1|<1, |\lambda_2|>1).$$\\[0.1in]
(iv)$(x*,y*)$ is a non-hyperbolic point if 
$$r\in[1+\sqrt{6},4)\mbox{ and } \epsilon =f_2(r)$$ (here $|\lambda_1|=|\lambda_2|=1, \lambda_{i}\in\mathbb{C}, i=1,2 $)
\mbox{ or }
$$r\in(3,4)\mbox{ and } \epsilon=\frac{1}{2}+\frac{\sqrt{3}}{2}\sqrt{\frac{1}{r(r-2)}} \mbox{ (specifically }, \lambda_2=-1, |\lambda_1|<1)$$ or \\
$$r\in(3,4)\mbox{ and } \epsilon=\frac{1}{2}-\frac{\sqrt{3}}{2}\sqrt{\frac{1}{r(r-2)}} \mbox{ (specifically }, \lambda_2=-1, |\lambda_1|>1)$$
or
 $r=3 \mbox{ and } \epsilon=0$ ,where our system now corresponds to an uncoupled pair of logistic maps.\\[0.1in]
{\bf Proof}.  For the trivial fixed point $(0,0)$, $\lambda_1=r,\mbox{  } \lambda_2=r(1-2\epsilon)$.  By inspection we see $|\lambda_{i}|<1, \mbox{ for i=1,2 if and only if } r\in(0,1) \mbox{ for any epsilon in } [0,1]$.  The remaining parts of A can easily be deduced. \\
For the symmetric fixed point $(\frac{r-1}{r},\frac{r-1}{r})$, $\lambda_1=r(1-2\epsilon)(\frac{2}{r}-1) \mbox{ and } \lambda_2=r(\frac{2}{r}-1)$.  Again a straightforward calculation shows that parts (i)-(iv) of B hold. \\[0.1in] 
For the anti-symmetric fixed point $(x*,y*)$ a direct calculation shows that the eigenvalues are $$ \lambda_{1}=\frac{{\epsilon-1}+(2\epsilon-1) \sqrt{\frac{\epsilon^2+(1-2\epsilon)\Delta}{(1-2\epsilon)^2)}}}{2\epsilon-1},\lambda_{2}=\frac{{\epsilon-1}+(1-2\epsilon) \sqrt{\frac{\epsilon^2+(1-2\epsilon)\Delta}{(1-2\epsilon)^2)}}}{2\epsilon-1}$$ from which one can establish (i)-(iv).\\[0.1in]
\begin{figure}[htp!]
\centering
\includegraphics[width=2.9in]{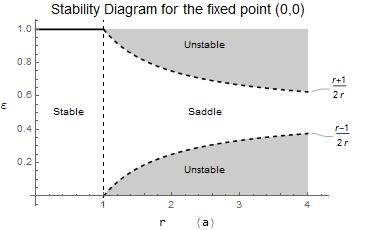}
\includegraphics[width=2.9in]{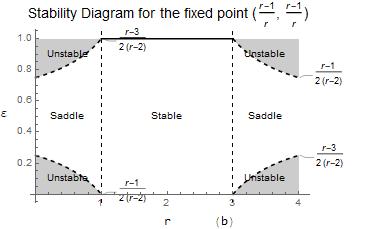}\\
\includegraphics[width=2.9in]{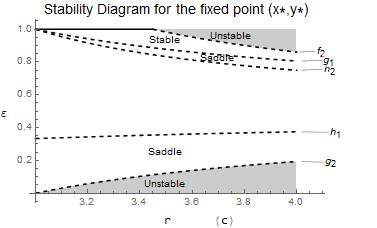}
\caption{Diagrams of the regions of stability for three of the four fixed points of system (\ref{main})  in the $(r,\epsilon)$ plane} \label{fig:series1}
\end{figure}\\
In Figure 1(a) we illustrate the stable, unstable and saddle regions for the fixed point $(0,0)$. Figures 1 (b) and (c) show these three regions for the fixed points $(\frac{r-1}{r},\frac{r-1}{r})$ and $(x*,y*)$. 

In 1(a) above the upper curve $\frac{r+1}{2r}$ is the flip curve, and $\frac{r-1}{2r}$, $r=1$ are fold curves.  In 1(b) the two upper dashed curves denote flip and fold curves respectively as well as the lines $r=3$ and $r=1$ respectively. In 1(c) we
define $h_1=\frac{r-1}{2r}, h_2=\frac{r-1}{2(r-1)},g_1=\frac{1}{2}+\frac{\sqrt{3}}{2}\sqrt{\frac{1}{r(r-2)}},g_2=\frac{1}{2}-\frac{\sqrt{3}}{2}\sqrt{\frac{1}{r(r-2)}}$ and $f_2=f_{2}(r)$ which was defined earlier. Here $g_1,g_2$ are flip curves, $f_{2}(r)$ is a Neimark-Sacker curve and $h_1,h_2$ are the curves bounding the saddle regions. We also note that for the two symmetric fixed points we have symmetric regions of stability/instability whose bounding curves the translation symmetry $\epsilon \mapsto 1-\epsilon$ inherent in the system's defining equations. For the anti-symmetric fixed point $(x*,y*)$ this translation symmetry manifests in the equations for the bounding curves $g_1,g_2$ but not in the regions bounded by these curves.\\

\section{Local Bifurcation Analysis}
\subsection{Flip Bifurcation}
Now we determine the stability of the non-hyperbolic fixed point $(x*,y*)$ via center manifold theory.
 In particular we demonstrate that system~(\ref{main}) undergoes a flip bifurcation at $(x*,y*)$ where $\lambda_{1}=-1$ and $ \lambda_{2}=\frac{4\epsilon-3}{2\epsilon-1}$ and where we choose $\epsilon$ as our bifurcation parameter and allow it to vary in a small neighborhood
 of $(x*,y*)$. Generically, a flip bifurcations is characterized by a 
 the loss of stability of a periodic orbit as a parameter crosses a critical value (from above or below) and at which point locally, either
 there exists stable periodic orbits with double the period for parameter values near the critical parameter forming a new branch that emerges at the critical parameter value (super-critical period doubling) or unstable periodic orbits with double the period coalescing with and  destroyed by stable periodic orbits (sub-critical period doubling). Moreover, a flip bifurcation occurs at an eigenvalue of -1 of the Jacobian of the map.\\ 
 In order to apply Center Manifold theory we assume that our discrete system has the form:
 \begin{equation*}
 x_{n+1}=Ax_{n}+F(x_{n},y_{n})
 \end{equation*}
   \begin{equation}
 y_{n+1}=By_{n}+F(x_{n},y_{n}) \label{main2}
 \end{equation}
where all of the eigenvalues of the matrix $A$ (an $n\times n$ matrix) are on the unit circle and the eigenvalues of the matrix $B$ (an $m x m$ matrix) are within the unit circle, and the Jacobian matrix for the system has the form 

$$\begin{bmatrix}
A & 0 \\
0 & B
\end{bmatrix}$$

We assume without loss of generality that the system has the origin as a fixed point.  We use a slight modification of the following version of the Center Manifold Theorem in$\mbox{ }$\cite{ELY}:\\[0.1in]
${\bf Theorem\mbox{ }5.1}$\mbox{  }
{\it There exists a $C^r$-center manifold for system ~(\ref{main2}) that can be represented locally as
$$W_{loc}^{c}(0,0)=\{(x,y,\mu) \in \mathbb{R}^3\mbox{ } |\mbox{ } y=h(x,\mu),|x|<\delta_1, |\mu|<\delta_2,,h(0,0)=D h(0,0), |x|<\epsilon,|\mu|<\delta\}$$
Furthermore, the dynamics of the system restricted to $W_{loc}^{c}(0)$ are given locally by the map $$ x\longmapsto Ax+ f(x,\mu,h(x,\mu)),\mbox{ for }x\in \mathbb{R}.$$}\\[0.1in]
In addition we state the following theorem from \cite{GH} which gives criteria for the existence of a flip bifurcation:\\[0.1in]
{\bf Theorem 3.5.1}\mbox{ }{\it Let $f_{\mu}:\mathbb{R}\rightarrow\mathbb{R}$ be a one parameter family of mappings such that $f_{\mu_{0}}$ has a fixed point $x_{0}$ with an eigenvalue of value $-1$.  Assume 
$$\frac{\partial f}{\partial \mu}\frac{\partial^2 f}{\partial x^2} + 2\frac{ \partial ^2 f}{\partial x \partial \mu}\neq 0 \mbox{ at } (x_{0},\mu_{0});$$
$$ \frac{1}{2}\left(\frac{\partial^2 f}{\partial x^2}\right)^2+\frac{1}{3}\left(\frac{\partial^3 f}{\partial x^3}\right)\neq 0 \mbox{ at } (x_{0},\mu_{0}).$$\\
Then there is a smooth curve of fixed points of $f_{\mu}$ passing through $(x_{0},\mu_{0})$, the stability of which changes at $(x_{0},\mu_{0})$.  There is also a smooth curve $\gamma$ passing through $(x_{0},\mu_{0})$ so that $\gamma-{(x_{0},\mu_{0})}$ is a union of hyperbolic period 2 orbits.  The curve $\gamma$ has quadratic tangency with the line $\mathbb{R} \times \{\mu_{0}\}$ at $(x_{0},\mu_{0})$.}

 We begin the establishment of a flip bifurcation at $(x*,y*)$ by first defining 
\begin{equation}
H_{FP}=\Bigg\{(r,\epsilon): r\in[3,4), \epsilon=\frac{1}{2}+\frac{\sqrt{3}}{2}\sqrt{\frac{1}{r(r-2)}}\Bigg\}
\end{equation}
 the set containing the parameters that satisfy the second condition for a hyperbolic point in C (iv) from Theorem 1. For arbitrary parameters $(r_{s},\epsilon_{s})\in H_{FP}$ 
and by the change of variables $u_{n}=x_{n}-x*,v_{n}=y_{n}-y*$ where we also set  $\bar{\epsilon}=\epsilon-\epsilon_s ( \mbox{ and so } \epsilon_s=\frac{1}{2}+\frac{\sqrt{3}}{2}\sqrt{\frac{1}{r(r-2)}})$ be a new independent variable, we transform the fixed point $(x*,y*)$ into $(0,0)$. System~(\ref{main}) now has the form \\[0.1in]
\begin{align}
  \left(  \begin{matrix}
           u_{n+1} \\
           v_{n+1}
           \end{matrix}\right)=
           \left(\begin{matrix}
           a_{11}u_n+a_{12}v_n+a_{13}u_{n}^2+a_{14}v_{n}^2+b^{*}\bar{\epsilon}+b_{13}\bar{\epsilon}u_{n}^2-b_{13}\bar{\epsilon}v_{n}^2\\
           a_{21}u_n+a_{22}v_n+a_{23}u_{n}^2+a_{24}v_{n}^2-b^{*}\bar{\epsilon}-b_{13}\bar{\epsilon}u_{n}^2+b_{13}\bar{\epsilon}v_{n}^2
           \end{matrix}\right)
                   \end{align}\\
    where 
   \begin{align}
    \begin{matrix}
    a_{11}=r_s(1-\epsilon_s)(1-2x*)& a_{21}=r_s\bar{\epsilon}(1-2x*)\\
    a_{12}=r_s\epsilon_s(1-2y*)& a_{22}=r_s(1-e_s)(1-2y*)\\
    a_{13}=r_s(\epsilon_s-1) & a_{14}=-r_s\epsilon_s
    \end{matrix}
    \end{align}
    and
    
    $$ b^{*}=r_s\bar{\epsilon}\left((x*)^2-x*+y*-(y*)^2-(1-2x*)u_{n}+(1-2y*)v_{n}\right),\mbox{ } b_{13}=r_s $$

  We begin the process of putting the system into the format of the equations in ~(\ref{main2}) by first defining an invertible matrix 
$$
T=\left(\begin{matrix}
-a_{12}&-a_{12}\\
a_{11}+1& a_{11}-\lambda_{2}\\
\end{matrix}\right)
$$
 
  determined by the eigenvectors associated with the linearization of the system at $(0,0)$.  Using the transformation 
  
 \begin{align}
 \left( \begin{matrix}
  u_{n}\\
  v_{n}
  \end{matrix}\right)=\mbox{ } T
 \left( \begin{matrix}
  X_{n}\\
  Y_{n}
  \end{matrix}\right)
  \end{align}
  and letting $\mu=\bar{\epsilon}$ the system now takes the desired form:
  \begin{align}
  \left(\begin{matrix}
  X_{n+1}\\
  Y_{n+1}
  \end{matrix}\right)=
  \left(\begin{matrix}
  -1 & 0\\
  0 & \lambda_{2}-a_{11}\\
  \end{matrix}\right)
  \left(\begin{matrix}
  X_{n}\\
  Y_{n}
  \end{matrix}\right)+
  \left(\begin{matrix}
  F(X_{n},Y_{n},\mu)\\
  G(X_{n},Y_{n},\mu)
  \end{matrix}\right)
  \end{align}
  where

$$F(X_n,Y_n,\mu) 
 = \frac{b_2}{a_{12}(1+\lambda_2)}\left[(a_{13}a_{12}^2-a_{13}b_{1})X^2_n+((a_{13}a_{12}^2-a_{13}b_{2})Y^2_n+(2a_{13}(a^2_{12}-b_1b_2)X_{n}Y_n+\mu b^*\right]$$\\
 $$ +\frac{b_2}{a_{12}(1+\lambda_2)}\left[b_{13}\mu([a^2_{12}-b^2_1]X_n^2+[a^2_{12}-b_2^2]Y_n^2+2[a^2_{12}-b_1b_2]X_{n}Y_{n})\right]$$\\
 $$ +\frac{1}{1+\lambda_2}\left[(a_{13}(b_1-a_{12}^2))X^2_n+(a_{13}(b_2-a_{12}^2))Y_n^2+(2a_{13}(b_1b_2-a_{12}^2)X_{n}Y_{n}-\mu b^*\right]$$\\
 $$ +\frac{1}{1+\lambda_2}\left[-b_{13}\mu([a_{12}^2-b_{1}^2]X^2_{n}+[a^2_{12}-b_{2}^2]Y^2_{n}+2[a^2_{12}-b_{1}b_{2}]X_{n}Y_{n}\right]$$

  and 
  $$G(X_n,Y_n,\mu)
 =\frac{-b_1}{a_{12}(1+\lambda_2)}\left[a_{13}(a_{12}^2-b_1)X_{n}^2+a_{13}(a_{12}^2-b_2)Y_{n}^2 +2(a_{13}(a_{12}^2-b_{1}b_{2}))X_{n}Y_{n}+\mu b^*\right]$$\\
 $$\frac{-b_1}{a_{12}(1+\lambda_2)}\left[b_{13} \mu\left[ (a_{12}^2-b_1^2)X_n^2+(a_{12}^2-b_2^2)Y_{n}^2+2(a_{12}^2-b_{1}b_{2})X_{n}Y_{n})X_{n}Y_{n}\right]\right]$$
 $$-\frac{1}{1+\lambda_2}\left[a_{13}(b_{1}-a_{12}^2)X_{n}^2+ a_{13}(b_2-a_{12}^2)Y_{n}^2+ 2a_{13}(b_{1}b_{2}-a_{12}^2)X_{n}Y_{n}-\mu b^*\right]$$
 $$-\frac{1}{1+\lambda_2}\left[-b_{13}\mu\left[(a_{12}^2-b_{1}^2)X_{n}^2+(a_{12}^2-b_{2}^2)Y_{n}^2+2(a_{12}^2-b_{1}b_{2})X_{n}Y_{n}\right]\right]$$
 where  $b_{1}=a_{11}+1, b_{2}=a_{11}-\lambda_{2}$.
By applying  the center manifold theorem we see that there exists a center manifold for system~(\ref{main}) defined as
$$W_{loc}^{c}(0,0)=\{(x,y,\mu) \in \mathbb{R^3}\mbox{ } |\mbox{ } y=h(x,\mu),|x|<\delta_1, |\mu|<\delta_2,,h(0,0)=D h(0,0), |x|<\epsilon,|\mu|<\delta\}$$ for sufficiently small $\epsilon \mbox{ and } \delta$. To actually find the center manifold as the graph of $y=h(x,\mu)$ we consider a power series representation for this map:
$$y=h(x,\mu)=A_{0}X^2+A_{1}X \mu +A_{2}\mu^2+ O((|X|+|\mu|)^3)$$ which we then substitute into~(\ref{main2}). Hence, the center manifold must satisfy the equation 
\begin{equation}
 h(-x+F(x,h(x,\mu),\mu),\mu)=\lambda_{2}h(x,\mu)+G(x,h(x,\mu),\mu).\label{cme}
 \end{equation}
By writing $F$ in the form
$$F(X,Y,\mu)=(f_{1}-g_{1})\left[e_{1}X^2+e_{2}Y^2+e_{3}X Y+\mu e_{4}+\mu e_{8}(e_{5}X^2+e_{6}Y^2+e_{7}X Y)\right]$$
and $G$ in the form
$$G(X,Y,\mu)=(f_{2}-g_{2})\left[e_{1}X^2+e_{2}Y^2+e_{3}X Y+\mu e_{4}+\mu e_{8}(e_{5}X^2+e_{6}Y^2+e_{7}X Y)\right]$$
where 
\begin{align}
    \begin{matrix}
    e_{1}=a_{13}(a_{12}^2-b_{1}) & e_{2}=a_{13}(a_{12}^2-b_{2}) \\
    e_{3}=2a_{13}(a_{12}^2-b_{1}b_{2}) & e_{4}=b^{*}\\
    e_{5}=a_{12}^2-b_{1}^2 & e_{6}=a_{12}^2-b_{2}^2\\
    e_{7}=2(a_{12}^2-b_{1}b_{2}) & e_{8}=b_{13}
    \end{matrix}
    \end{align}\\

$$f_{1}=\frac{b_2}{a_{12}(1+\lambda_{2})}, g_{1}=\frac{1}{1+\lambda_2},g_{2}=-g_{1},f_{2}=\frac{-b_1}{a_{12}(1+\lambda_{2})}$$\\
By substituting the equations for $F$, $G$ and $h$ into the center manifold equation  ~(\ref{cme}) and equating the coefficients of like terms on either side of the equation, we determine the coefficients $A_{0}, A_{1}, A_{2}:$\\
$$A_{0}=\frac{(f_{2}-g_{2})e_{1}}{1-\lambda_{2}} $$
$$A_{1}=\frac{2A_{0}(g_{1}-f_{1})\mu e_{4}}{1+\lambda_2}$$
$$A_{2}=\frac{(f_{1}-g_{1})[A_{0}e_{4}^2+A_{1}e_{4}]}{\lambda_{2}-1} $$\\

The restriction of our map to the center manifold is defined as the map
$$K(X,\mu):= -X +(f_{1}-g_{1})\left[(e_{1}+\mu e_{8}e_{5})X^2+(e_{3}+\mu e_{8}e_{7})(X^{3}+A_{1}X^2 \mu +A_{2}\mu ^2 X)\right]$$
$$+(f_{1}-g_{1})\left[(e_{2}+\mu e_{8}e_{6})(A_{0}^2X^{4}+A_{1}^2X^2\mu^2+A_{2}^2\mu^4 +2A_{0}A_{2}X^2\mu^2+2A_{0}X^3A_{1}\mu+2A_{1}A_{2}\mu^4 X\right].$$\\[0.1in]
Straightforward but detailed calculations shows that 
$$\alpha_1=\frac{\partial K}{\partial \mu}\frac{\partial^2 K}{\partial X^2} + \frac{2 \partial ^2 K}{\partial X \partial \mu}=2e_{1}e_{4}(f_{1}-g_{1})^2 \Bigr\rvert_{(0,0)}\neq{ 0} $$\\
and\\
$$\alpha_2=\frac{1}{2}\left(\frac{\partial^2 K}{\partial X^2}\right)^2+\frac{1}{3}\left(\frac{\partial^3 K}{\partial X^3}\right)=2(f_{1}-g_{1})[(f_{1}-g_{1})e_{1}^2+A_{0}e_{3}]\Bigr\rvert_{(0,0)}\neq 0 $$\\[0.1in] 
By Theorem 5.1 and Theorem 3.5.1 above, the following result is now established:\\[0.2in]
{\bf Theorem 2}\mbox{ }
{\it If $\alpha_1,\alpha_2 \neq 0$ then the map undergoes a flip bifurcation at the fixed point $(x*,y*)$ when the parameter $\epsilon$ varies in a small neighborhood of $\epsilon_{s}$.  Moreover if $\alpha_{2}>0$ ( respectively $ \alpha_{2} <0)$ the period-2 orbits that bifurcate from $(x*,y*)$ are stable (unstable).}\\[0.1in]

\subsection{Neimark-Sacker Bifurcation}
A Neimark-Sacker bifurcation is characterized by a stable fixed point becoming unstable at a certain critical value of the bifurcation parameter of the system in which an an attracting closed invariant curve manifests or a repelling closed invariant curve emerges as the values of the parameter cross this critical value.
In the former case, we say the bifurcation is a supercritical Neimark -Sacker bifurcation; in the latter case a subcritical Neimark-Sacker bifurcation. In either case such a a bifurcation is associated with discrete systems whose eigenvalues are complex conjugates of modulus one.\\

Here we state a slight modification of a theorem from \cite{ELY}, (Chapter 5), which outlines the criteria for the emergence of such a bifurcation.\\[0.1in]
 {\bf Theorem 5.4 (Neimark-Sacker)\mbox{ }}
 {\it Consider the family of $C^{r}$ maps $(r\geq 5), F_{\mu}:\mathbb{R^2} \times \mathbb{R}\rightarrow\mathbb{R^2}$ such that the following conditions hold:\\
 \begin{align*}
 &1.\mbox{ } F_{\mu}(0)=0, \mbox{i.e., the origin is a fixed point of } F_{\mu}.\\
 &2.\mbox{ } DF_{\mu}(0)\mbox{ has two complex conjugate eigenvalues } \lambda_{1,2}(\mu)=r(\mu)e^{\pm{i\theta(\mu)}},
  \mbox{ where } r(0)=1, r'(0)\neq 0, \theta(0)=\theta_{0}.\\
 &3.\mbox{ } e^{ik\theta_{0}}\neq 1 \mbox{ for }k=1,2,3,4\mbox{ (absence of strong resonances condition)}.
 \end{align*}
 
If in addition, $a\neq 0$ where
 \begin{align*}
    a=-Re\left[\frac{(1-2\lambda)\bar{\lambda}^2\zeta_{11}\zeta_{20}}{1-\lambda}\right]-\frac{1}{2}|\zeta_{11}|^2-|\zeta_{02}|^2+Re(\bar{\lambda}\zeta_{21})\mbox{, } (a \mbox{ is called the first Lyapunov coefficient}),
\end{align*}
then for sufficiently small $\mu$, $F_{\mu}$ there exists a unique invariant closed curve enclosing that bifurcates from the origin as a passes through 0. If a>0 we have a supercritical Neimark-Sacker bifurcation.  If $a<0$ we have a subcritical Neimark-Sacker bifurcation.}\\[0.1in]
The complex conjugate eigenvalues of our system are given by the following formulas:
 \begin{equation}
    \lambda_{i}=\frac{{\epsilon-1}\pm i(2\epsilon-1) \sqrt{\frac{-\epsilon^2+(2\epsilon-1)\Delta}{(1-2\epsilon)^2)}}}{2\epsilon-1}, \mbox{ for } i=1,2.
\end{equation}
A simple calculation shows that 
\begin{equation}
    |\lambda_i|=1\mbox{, for i=1,2} \mbox{ if and only if} \sqrt{\frac{\Delta -1}{2\epsilon-1}}=1 , \mbox{ or }
    \Delta=2\epsilon.
\end{equation}
Thus the range of parameters for which the eigenvalues associated with the fixed point $(x*,y*)$ are complex conjugates and have magnitude 1 can be described by the set
\begin{equation}
    H_{NS}=\{(r,\epsilon): \epsilon^2+(1-2\epsilon)\Delta<0, \Delta=2\epsilon\}\equiv\{(r,\epsilon): r\in[1+\sqrt{6},4),\epsilon=f_{2}(r)\}
\end{equation}

We now show that a Neimark-Sacker bifurcation occurs at $(x*,y*)$ for arbitrary parameters $(e_h,r_h)\in H_{NS}$, taking $\epsilon$ as our bifurcation parameter and allowing it to vary in a small neighborhood of $e_h$.
So we consider a small perturbation of the parameter $\epsilon$ as follows: $\bar{\bar{\epsilon}}=\epsilon-\epsilon_h$ and transform the fixed point $(x*,y*)$ to the origin $(0,0)$ as before to produce the system (where we are essentially
replacing $e_{s}$ by $e_h$ in an earlier statement of our system) with coefficients that were defined in Section 3:
\begin{align}
  \left(  \begin{matrix}
           u_{n+1} \\
           v_{n+1}
           \end{matrix}\right)=
           \left(\begin{matrix}
           a_{11}u_n+a_{12}v_n+a_{13}u_{n}^2+a_{14}v_{n}^2+b^{*}\bar{\bar{\epsilon}}+b_{13}\bar{\bar{\epsilon}}u_{n}^2-b_{13}\bar{\bar{\epsilon}}v_{n}^2\\
           a_{21}u_n+a_{22}v_n+a_{23}u_{n}^2+a_{24}v_{n}^2-b^{*}\bar{\bar{\epsilon}}-b_{13}\bar{\bar{\epsilon}}u_{n}^2+b_{13}\bar{\bar{\epsilon}}v_{n}^2
           \end{matrix}\right)
                   \end{align}\\
Now the characteristic equation at $(u_n,v_n)=(0,0)$ is as follows:
\begin{equation}
\lambda^2-\lambda(r_h(1-\epsilon_{h}-\bar{\bar{\epsilon}})(1-2x*)+r_h(\epsilon_{h}+\bar{\bar{\epsilon}})(1-2y*))+r_h^2 (1-2(\epsilon_{h}-\bar{\bar{\epsilon}}))(1-2x*)(1-2y*)
\end{equation}
where 
 
\begin{equation}
    \lambda_{1,2}=\frac{{(\epsilon_h+\bar{\bar{\epsilon}})-1}\pm i(2(\epsilon_h+\bar{\bar{\epsilon}})-1) \sqrt{\frac{-(\epsilon_h+\bar{\bar{\epsilon}})^2+(2(\epsilon_h+\bar{\bar{\epsilon}})-1)\Delta}{(1-2( \epsilon_h+\bar{\bar{\epsilon}}))^2)}}}{2(\epsilon_h+\bar{\bar{\epsilon}})-1}.
\end{equation}

A straightforward calculation shows that
\begin{equation}
    \frac{d}{d\bar{\bar{\epsilon}}}(|\lambda_{1,2}|)= \frac{d}{d\bar{\bar{\epsilon}}}(\sqrt{\frac{\Tilde\Delta -1}{(2(\bar{\bar{\epsilon}}+\epsilon_h)-1)})}\biggr\rvert_{\bar{\bar{\epsilon}}=0} = \frac{2}{(1-2\epsilon_h)^2}+(r_h)^2-2r_h >0 \mbox{ for } (r_h,\epsilon _h)\in H_{NS}, \mbox{ where }
    \end{equation}
    \begin{equation*}
    \Tilde\Delta=(1-4(\bar{\bar{\epsilon}}+\epsilon_h))(r-1)^2+4(\bar{\bar{\epsilon}}+\epsilon_h^2)r_h(r_h-2)2(\bar\bar{{\epsilon}}+\epsilon_h).
\end{equation*}\\[0.1in]
Now we state conditions for the absence of strong resonances, i.e.
$\lambda^m_{1,2}(\epsilon_h)\neq 1, m=1,2,3,4$ for $\bar{\bar{\epsilon}}=0$.  Here we note that the condition that the eigenvalues are a pair of complex conjugates leads to the following condition deducible from equation (17) using $\Delta =2\epsilon$:
We can write 
\begin{equation}
    \lambda_{1,2}=\frac{\epsilon-1\pm i(2\epsilon-1)\sqrt{\frac{3\epsilon^2-2\epsilon}{(1-2\epsilon)^2}}}{2\epsilon-1}
\end{equation}
An examination of the condition $\lambda^m (e_h)\neq 1$ for $m=1,2,3,4$, leads to the constraints
$\epsilon\neq 0,\frac{2}{3},\frac{3}{4},1$.  For $r\in[1+\sqrt{6},4)$ these $\epsilon$ constraints, again for $\epsilon \in H_{NS}$, are equivalent to $r \neq 1+\sqrt{6}$ which we now require.
Now we study the normal form of our system when $\bar{\bar{\epsilon}}=0$ by first computing the following Taylor expansion at $(u_n,v_n)=(0,0)$:
\begin{align}
  \left(  \begin{matrix}
           u_{n+1} \\
           v_{n+1}
           \end{matrix}\right)=
           \left(\begin{matrix}
           a_{11}u_n+a_{12}v_n+a_{13}u_{n}^2+a_{14}v_{n}^2\\
           a_{21}u_n+a_{22}v_n+a_{14}u_{n}^2+a_{13}v_{n}^2
           \end{matrix}\right)
                   \end{align}\\
where the coefficients $a_{11},a_{21},a_{12},a_{13},a_{14},a_{22}$ were defined earlier.
Next we define $A_{1}=\frac{\epsilon-1}{2\epsilon-1}$ and $A_{2}=\sqrt{\frac{3\epsilon^2-2\epsilon}{(1-2\epsilon)^2}}$; these coefficients represent the real and imaginary parts of $\lambda_{1,2}$. Upon finding the eigenvectors associated with these eigenvalues we
construct the following invertible matrix 
$$
T=\left(\begin{matrix}
-a_{12}& 0\\
a_{11}-A_{1}& A_{2}\\
\end{matrix}\right)
$$
Using the transformation
\begin{align}
 \left( \begin{matrix}
  u_{n}\\
  v_{n}
  \end{matrix}\right)=\mbox{ } T
 \left( \begin{matrix}
  X_{n}\\
  Y_{n}
  \end{matrix}\right)
  \end{align}
the system can be rendered in the form
\begin{equation}
    X_{n+1}=A_{1}X_{n}-A_{2}Y_{n}+F(X_{n},Y_{n})
\end{equation}
\begin{equation}
    Y_{n+1}=A_{3}X_{n}+A_{4}Y_{n}+G(X_{n},Y_{n})
\end{equation}
where \begin{equation}
    F(X_{n},Y_{n})=c_{11}X_{n}^2+c_{12}X_{n}Y_{n}+c_{13}Y_{n}^2
\end{equation}
and \begin{equation}
    G(X_{n},Y_{n})=c_{21}X_{n}^2+c_{22}X_{n}Y_{n}+c_{23}Y_{n}^2
\end{equation}
Here, the coefficients are defined as 
\begin{align}
    A_{3}=\frac{A_{1}^2-A_{1}(a_{11}+a_{22})+a_{11}a_{22}-a_{21}a_{12}}{A_{2}}, 
 \end{align}
 \begin{align}
       A_{4}=a_{11}+a_{22}-A_{1}
 \end{align}
\begin{align}
    c_{21}=\frac{A_{1}^2a_{13}-2A_{1}a_{11}a_{13}+(a_{11})^2a_{13}-A_{1}a_{12}a_{23}+a_{11}a_{12}a_{23}+a_{12}^2a_{23}}{A_{2}}\\
    +\frac{a_{11}^3a_{13}+3A_{1}^2a_{11}a_{13}-A_{1}^3a_{13}-3A_{1}a_{11}^2a_{13}}{a_{12}A_{2}}
\end{align}
\begin{equation}
    c_{22}=2a_{11}a_{13}-2A_{1}a_{13}+\frac{2A_{1}^2a_{13}-4A_{1}a_{11}a_{13}+2a_{11}^2a_{13}}{a_{12}}
\end{equation}
\begin{equation}
    c_{23}=A_{2}a_{13}+\frac{-A_{1}A_{2}a_{13}+a_{11}A_{2}a_{13}}{a_{12}}
\end{equation}
\begin{equation}
   c_{11}=\frac{2A_{1}a_{11}a_{23}-A_{1}^2 a_{13}-a_{11}^2 a_{23}}{a_{12}}
\end{equation}

\begin{equation}
c_{12}=\frac{2A_{1}A_{2}a_{23}-2a_{11}A_{2}a_{23}}{a_{12}}
\end{equation}
\begin{equation}
   c_{13}= \frac{A_{2}^2a_{23}}{a_{12}}
\end{equation}
In addition we have
\begin{gather*}
   F_{x_{n}x_{n}} \Bigr\rvert_{(0,0)}=2c_{11}\\
   F_{x_{n}y_{n}}\Bigr\rvert_{(0,0)}=c_{12}\\
   F_{y_{n}y_{n}}\Bigr\rvert_{(0,0)}=2c_{13}\\
   F_{x_{n}x_{n}x_{n}}\Bigr\rvert_{(0,0)}=F_{x_{n}x_{n}y_{n}}\Bigr\rvert_{(0,0)}=F_{x_{n}y_{n}y_{n}}\Bigr\rvert_{(0,0)}=F_{y_{n}y_{n}y_{n}}\Bigr\rvert_{(0,0)}=0
\end{gather*}
and 
\begin{gather*}
    G_{x_{n}x_{n}}\Bigr\rvert_{(0,0)}=2c_{21}\\
    G_{x_{n}y_{n}}\Bigr\rvert_{(0,0)}=c_{22}\\
    G_{y_{n}y_{n}}\Bigr\rvert_{(0,0)}=2c_{23}\\
     G_{x_{n}x_{n}x_{n}}\Bigr\rvert_{(0,0)}=G_{x_{n}x_{n}y_{n}}\Bigr\rvert_{(0,0)}=G_{x_{n}y_{n}y_{n}}\Bigr\rvert_{(0,0)}=G_{y_{n}y_{n}y_{n}}\Bigr\rvert_{(0,0)}=0
\end{gather*}
Now we must show that $a\neq 0$ where $\lambda,\bar{\lambda}=e^{\pm i\theta}$ and
\begin{equation}
    a=-Re\left[\frac{(1-2\lambda)\bar{\lambda}^2\zeta_{11}\zeta_{20}}{1-\lambda}\right]-\frac{1}{2}|\zeta_{11}|^2-|\zeta_{02}|^2+Re(\bar{\lambda}\zeta_{21})
\end{equation}
where
\begin{gather*}
    \zeta_{20}=\frac{1}{8}\left[(F_{x_{n}x_{n}}-F_{y_{n}y_{n}}+2G_{x_{n}y_{n}})+ i (G_{x_{n}x_{n}}-G_{y_{n}y_{n}}-2F_{x_{n}y_{n}}\right]\Bigr\rvert_{(0,0)}=\frac{1}{4}\left[(c_{11}-c_{13}+c_{22})+i(c_{21}-c_{23}-c_{12})\right]\\
    \zeta_{11}=\frac{1}{4}\left[(F_{x_{n}x_{n}}+F_{y_{n}y_{n}})+i(G_{x_{n}x_{n}}+G_{y_{n}y_{n}}\right]\Bigr\rvert_{(0,0)}=\frac{1}{2}\left[(c_{11}+c_{13}+i(c_{21}+c_{23})\right]\\
    \zeta_{02}=\frac{1}{8}\left[(F_{x_{n}x_{n}}-F_{y_{n}y_{n}}-2G_{x_{n}y_{n}})+ i (G_{x_{n}x_{n}}-G_{y_{n}y_{n}}+2F_{x_{n}y_{n}}\right]\Bigr\rvert_{(0,0)}=\frac{1}{4}\left[(c_{11}-c_{13}-c_{22})+i(c_{21}-c_{23}+c_{12})\right]\\
    \zeta_{21}=\frac{1}{16}\left[( F_{x_{n}x_{n}x_{n}}+ F_{x_{n}y_{n}y_{n}}+ G_{x_{n}x_{n}y_{n}}+ G_{y_{n}y_{n}y_{n}})+ i( G_{x_{n}x_{n}x_{n}}+ G_{x_{n}y_{n}y_{n}}- F_{x_{n}x_{n}y_{n}}- F_{y_{n}y_{n}y_{n}}\right]\Bigr\rvert_{(0,0)}=0
\end{gather*}\\[0.1in]
We summarize our work now as a theorem indicating that a Neimark-Sacker bifurcation occurs at $(x*,y*)$ and the nature of the resulting bifurcation curve: \\[0.1in]
{\bf Theorem 3} If $r\neq 1+\sqrt{6}$ and $a\neq 0$ then the map undergoes a Neimark-Sacker bifurcation at the fixed point $(x*,y*)$ when the parameter $\epsilon$ varies in a small neighborhood of $\epsilon_h$.  Moreover if $a<0$ (respectively $a>0$ ) then an attracting (respectively repelling) invariant closed curve bifurcates from the fixed point for $\epsilon>\epsilon_h$ (respectively $\epsilon < \epsilon_h$).

\section{Numerical Results}
In this section we use Mathematica to numerically verify and illustrate the conclusions of Theorems 1, 2 and 3 with respect to the fixed point $(x*,y*)$. 
\begin{center}
\includegraphics[width=3.1in]{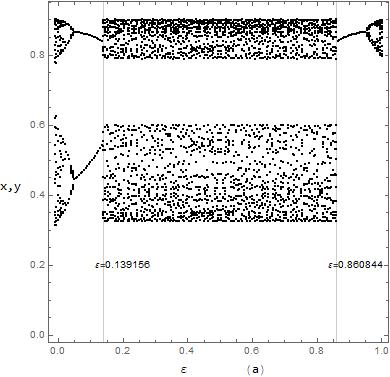}\\
\captionof{figure}{Bifurcation diagram for $r=3.6, \epsilon=0.860844$.}
\end{center}

Using the flip equation $\epsilon=\frac{1}{2}+\frac{\sqrt{3}}{2}\sqrt{\frac{1}{r(r-2)}}$ for $r=3.6$ we have $\epsilon=0.860844$ and $(x*,y*)=(.548868,.836032)$ and $\alpha_2=-15.6546$. Since the corresponding value $\alpha_2$<0 the period-2 orbits that bifurcate from $(x*,y*)$ are unstable and they are succeeded by a stable period-1 orbit. In figure 2, we observe the emergence of the period-1 orbit at the bifurcation point. The flip bifurcation occurs at $\epsilon=0.860844$. Here we include a vertical line at $\epsilon=0.139156$ to show at least numerically that there is another flip bifurcation for $\epsilon=\frac{1}{2}-\frac{\sqrt{3}}{2}\sqrt{\frac{1}{r(r-2)}}$. Figure 3 shows that the unstable flip occurs in the chaotic region and the subsequent stable one cycle thereafter. 
\begin{figure}[htp]
\centering
\includegraphics[width=3.1in]{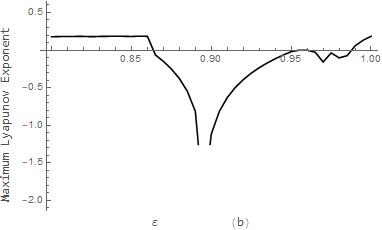}
\caption*{Figure 3:  Maximum Lyapunov Exponent Plot for $r=3.6$.} 
\end{figure}\\

In Figures 4, 5, and 6 we show further numerical evidence of a flip bifurcation at several other values of $r$.  Next we consider $r=3.1$ which corresponds to $\epsilon=0.968979$ . Here the corresponding fixed point is $(0.611386,0.732523)$ and the value of $\alpha_2=-0.225324$. The bifurcation diagram in Figure 4 shows the onset of flip bifurcations at the two marked off vertical lines $\epsilon=0.031021,\epsilon=0.968979 $.
\begin{figure}[htp]\
\centering
\includegraphics[width=3.1in]{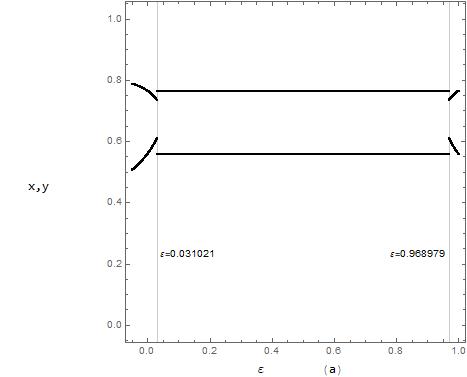}
\caption*{Figure 4: Bifurcation diagram for $r=3.1$, $\epsilon=0.968979$,with initial conditions $(0.09,0.18)$.}
\end{figure}

Figure 5 gives a sequence of time series plots revealing a stable symmetric two cycle before the critical value of $\epsilon$ is reached and a weak two cycle at the critical value of $\epsilon$. In the last plot we see the emergence of a one cycle for a value of $\epsilon$ nearby but larger than our critical value.  Here the chosen values of $\epsilon$ are $0.95,0.968979,0.988979$ respectively. \\
\begin{figure}[htp]\
\centering
\includegraphics[width=3.1in]{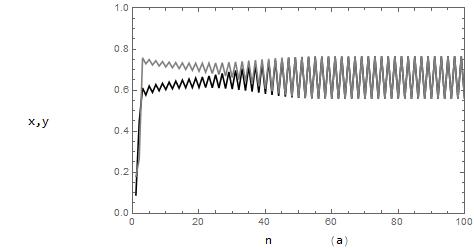}
\includegraphics[width=3.1in]{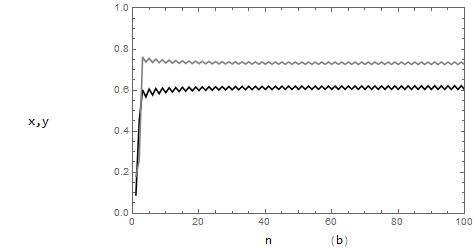}\\
\includegraphics[width=3.1in]{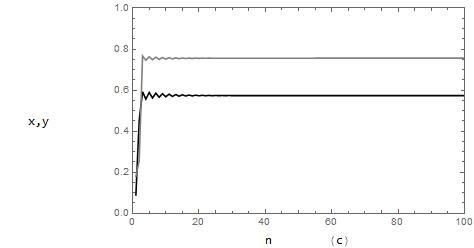}
\caption*{Figure 5: Time Series plots for $r=3.1$ , $\epsilon=0.95,0.968979,0.988979$ respectively.}
\end{figure}\\
For contrast, we consider a fairly high value of $r=3.83$, deep into the chaotic regime of the system. Here $\epsilon=0.827119$ and the initial conditions are $(0.24,0.7)$. The fixed point is $(0.533607,0.865478)$ and $\alpha_2=-38.6552$ . (The corresponding lower value of $\epsilon$ where a flip may occur is $\epsilon=0.172881$). The accompanying sequence of time series plots shows a chaotic cycle colliding with a two cycle at our critical value and the birth of a one cycle for a value of $\epsilon>0.827119$ close to our critical value. Additional time series plots (not included here) in fact show a pattern of intermittency-periods of stability and instability of a symmetric and anti-symmetric two cycles- before the one cycle is reached. In the panel the chosen values of $\epsilon$ are $0.807119,.827119,0.84,0.867$ respectively.\\
\begin{figure}[htp]\
\centering
\includegraphics[width=3.1in]{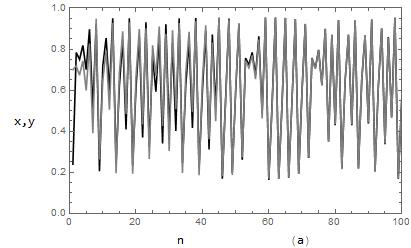}
\includegraphics[width=3.1in]{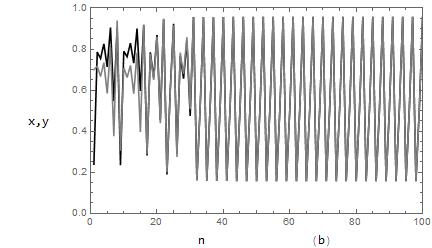}\\
\includegraphics[width=3.1in]{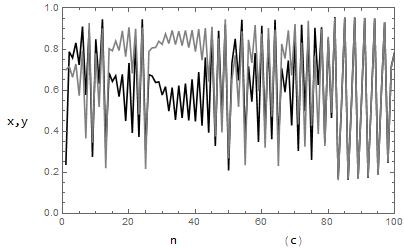}
\includegraphics[width=3.1in]{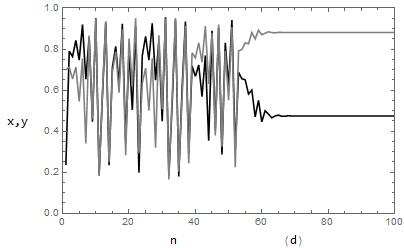}\\
\caption*{Figure 6:  Time Series plots for $r=3.83$ , $\epsilon=0.807119,.827119,0.84,0.867$,respectively.}
\end{figure}

Using the relation $\epsilon=f_{2}(r)$ and substituting $3.94$ for $r$ we get that $\epsilon=0.872059$ and the fixed point $(x*,y*)=(0.445316,0.895769)$.  Figures 7 and 8 below show the formation
of a Neimark-Sacker bifurcation and chaotic regions in the phase plane for the initial conditions $(0.1,0.3)$.
\begin{figure}[htp]
\centering
\includegraphics[width=2.9in]{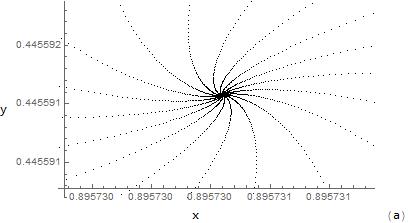}
\includegraphics[width=2.9in]{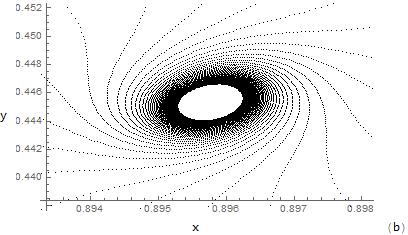}\\
\end{figure}
In Figure 7(a)  where $\epsilon=0.8718<0.872059 $ the fixed point is stable. Figures 7(b) illustrates the loss of stability of the fixed point at $\epsilon=0.872059$.    
\begin{figure}[htp]
\centering
\includegraphics[width=3.1in]{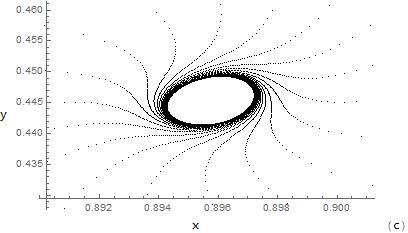}
\includegraphics[width=3.1in]{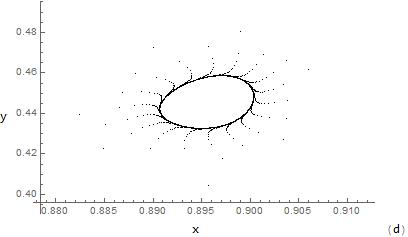}\\
\includegraphics[width=3.1in]{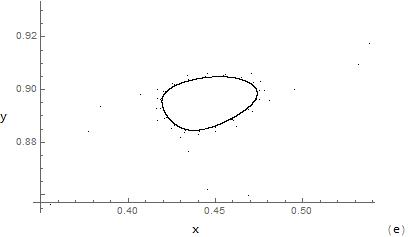}
\includegraphics[width=3.1in]{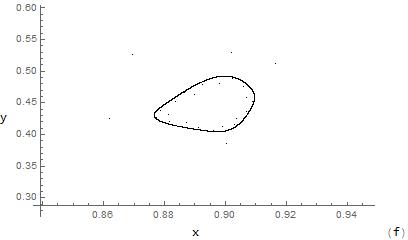}\\
\caption*{Figure 7: Formation of a Neimark-Sacker Bifurcation}
\end{figure}
In figures 7(c),(d),(e) and (f) $\epsilon=0.8721,0.8725,0.874,0.877$, respectively.

Here we see that for increasing $\epsilon>$0.872059 relatively close to $\epsilon=0.872059$ the gradual development of a closed invariant curve, in other words, a subcritical Neimark-Sacker bifurcation occurs. In addition, A detailed computation of $a$ yields a negative value.  Furthermore, in figures, 8(a),8(b), 8(c) and 8(d) (here $\epsilon=0.885,0.888,0.89,0.92)$ show the transition to a chaotic state with the appearance of 11 coexisting chaotic attractors in figure 8(b) and a chaotic attracting set in figures 8(c) and 8(d) , for values of $\epsilon$ further away from $0.872059$.

\begin{center}
\includegraphics[width=3.1in]{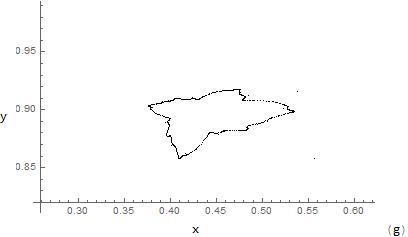}
\includegraphics[width=3.1in]{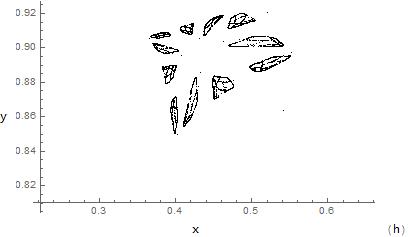}\\
\includegraphics[width=3.1in]{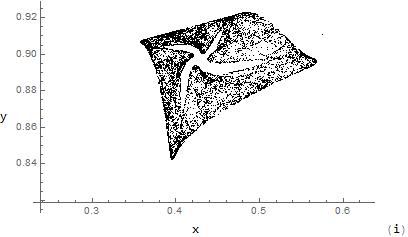}
\includegraphics[width=3.1in]{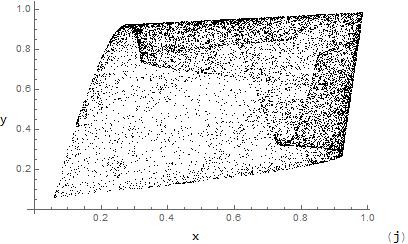}\\
\captionof*{figure}{Figure 8: Emergence of chaos} \label{series2}
\end{center}

The vertical line in the accompanying bifurcation diagram shows the birth of Neimark-Sacker bifurcation.  A plot of the maximum Lyapunov exponent for $r=3.94$ for $\epsilon$ in the range $[0.8,1]$ is also included.
\begin{figure}[htp]
\centering
\includegraphics[width=2.9in]{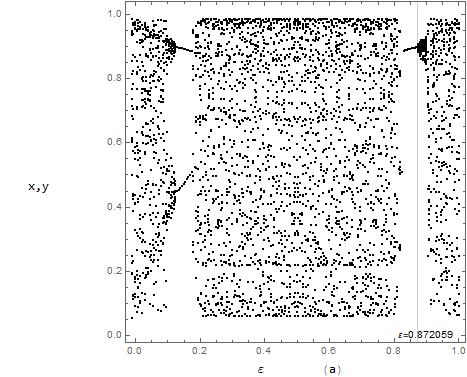}\\
\includegraphics[width=2.9in]{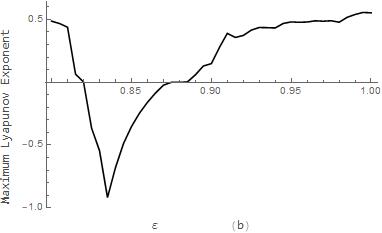}
\caption*{Figure 9: Bifurcation diagram and plot of the maximum Lyapunov exponent for $r=3.94$.}\label{series3}
\end{figure}
Negative exponents indicate stable regions within the otherwise chaotic regime and positive exponents are indicative of the chaotic regions.   
\section{Conclusion}
In this work we investigated the dynamics of a discrete coupled system of logistic maps. We determined the stability of the systems' fixed points and used center manifold and bifurcation theory to prove the existence of a flip and Neimark-Sacker bifurcation for the non-symmetric fixed point $(x*,y*)$.  Using $\epsilon$ as our bifurcation parameter our numerical results revealed that the flip bifurcation is a reverse flip bifurcation (or period halving bifurcation) in that at the critical value of the parameter a newly unstable period 2 cycles bifurcates to a stable period 1 cycle (rather than a 1 cycle becoming unstable and giving rise to a stable period 2 cycle). This result contrasts the usual 'period doubling cascade' observed in logistic map systems where typically $r$ (not $\epsilon)$ is chosen to be the bifurcation parameter.  A general examination of the constant $a$ in Theorem 3 and our numerical evidence show that the Neimark- Sacker bifurcation is subcritical.  The rich dynamics of the system also includes interesting chaotic sets which will be analyzed further in a forthcoming work.





\end{document}